\documentclass[conference]{IEEEtran}

\usepackage{amsmath}
\usepackage{amssymb}
\usepackage[dvips]{graphicx}
\usepackage{epsfig}

\newcommand{\K}{{\sf{K}}}

\newcommand{\E}{\mathcal{E}}

\newcommand{\C}{{\sf{C}}}

\newcommand{\tsnr}{{\text{\footnotesize{SNR}}}}

\newcommand{\ssnr}{\text{\scriptsize{SNR}}}

\newtheorem{theo:awgnderivatives}{Theorem}
\newtheorem{theo:cfderivatives}[theo:awgnderivatives]{Theorem}
\newtheorem{theo:ncfderivatives}[theo:awgnderivatives]{Theorem}

\newtheorem{corr:awgnasympt}{Corollary}
\newtheorem{corr:awgnbitenergy}[corr:awgnasympt]{Corollary}
\newtheorem{corr:awgnM3bitenergy}[corr:awgnasympt]{Corollary}
\newtheorem{corr:ncfadingasympt}[corr:awgnasympt]{Corollary}
\newtheorem{corr:ncfadingbitenergy}[corr:awgnasympt]{Corollary}

\begin{document}

\title{On the Low-SNR Capacity of Phase-Shift Keying with Hard-Decision Detection}



%
\author{\authorblockN{Mustafa Cenk Gursoy}
\authorblockA{Department of Electrical Engineering\\
University of Nebraska-Lincoln, Lincoln, NE 68588\\ Email:
gursoy@engr.unl.edu}}


\maketitle

\begin{abstract}\footnote{This work was supported in part by the NSF CAREER Grant
CCF-0546384.} The low-$\tsnr$ capacity of $M$-ary PSK transmission
over both the additive white Gaussian noise (AWGN) and fading
channels is analyzed when hard-decision detection is employed at the
receiver. Closed-form expressions for the first and second
derivatives of the capacity at zero $\tsnr$ are obtained. The
spectral-efficiency/bit-energy tradeoff in the low-$\tsnr$ regime is
analyzed by finding the wideband slope and the bit energy required
at zero spectral efficiency. Practical design guidelines are drawn
from the information-theoretic analysis. The fading channel analysis
is conducted for both coherent and noncoherent cases, and the
performance penalty in the low-power regime for not knowing the
channel is identified.
\end{abstract}

\section{Introduction}

Phase modulation is a widely used technique for information
transmission, and the performance of coded phase modulation has been
of interest in the research community since the 1960s. One of the
early works was conducted in \cite{Wyner} where the capacity and
error exponents of a continuous-phase modulated system, in which the
transmitted phase can assume any value in $[-\pi,\pi)$, is studied.
More recent studies include \cite{Pierce}, \cite{Kramer},
\cite{Kaplan}, \cite{Peleg}, \cite{Gursoy-part2}, and \cite{Zhang}.
Kaplan and Shamai studied in \cite{Kaplan} the achievable
information rates of differential phase-shift keying (DPSK) while
reference \cite{Peleg} investigated the capacity of $M$-ary PSK over
an additive white Gaussian noise (AWGN) channel with unknown phase
that stays constant for a block of symbols. Pierce in \cite{Pierce}
considered hard-decision detection of PSK signals trasmitted over
the AWGN channel and compared the performances of 2-, 3- and 4-phase
modulations. Pierce also provided in \cite{Pierce} an expression for
the bit energy required by $M$-ary PSK at zero spectral efficiency.
The authors in \cite{Kramer} analyzed the spectral efficiency of
coded PSK and DPSK with soft- and hard-decision detection. Reference
\cite{Gursoy-part2} analyzed the energy efficiency of PSK when it is
combined with on-off keying for transmission over noncoherent Rician
fading channels. Recent work by Zhang and Laneman \cite{Zhang}
investigated the achievable rates of PSK over noncoherent Rayleigh
fading channels with memory.

The low-SNR capacity of PSK with soft detection is well-understood.
For instance, Verd\'u \cite{Verdu} has shown that quaternary PSK
(QPSK) transmission over the AWGN or coherent fading channels is
optimally efficient in the low-SNR regime, achieving both the
minimum bit energy of $-1.59$ dB and optimal wideband slope which is
defined as the slope of the spectral efficiency curve at zero
spectral efficiency. Although soft detection gives the best
performance, hard-decision detection and decoding is preferred when
reduction in the computational burden is required \cite{Proakis}.
Such a requirement, for instance, may be enforced in sensor networks
\cite{Luo}. Moreover, at very high transmission rates such as in
fiber optic communications, obtaining multiple-bit resolution from
A/D converters may not be possible \cite{Kramer}. Finally, it is of
interest to understand the fundamental limits of hard-decision
detection so that the performance gains of soft detection can be
identified and weighed with its increased complexity requirements.
Motivated by these considerations and the fact that the performance
difference of hard and soft detections are more emphasized at low
power levels, we study in this paper the low-SNR capacity of $M$-ary
PSK over both the AWGN and fading channels when a hard-decision
detection is employed at the receiver end.

\section{Channel Model} \label{sec:channelmodel}

We consider the following channel model
\begin{gather}
r_k = h_ks_{x_k} + n_k \quad k = 1,2,3 \ldots
\end{gather}
where $x_k$ is the discrete input, $s_{x_k}$ is the transmitted
signal when the input is $x_k$, and $r_k$ is the received signal
during the the $k^{\text{th}}$ symbol duration. $h_k$ is the channel
gain. $h_k$ is a fixed constant in unfaded AWGN channels, while in
flat fading channels, $h_k$ denotes the fading coefficient.
$\{n_k\}$ is a sequence of independent and identically distributed
(i.i.d.) zero-mean circularly symmetric Gaussian random variables
denoting the additive background noise. The variance of $n_k$ is
$E\{|n_k|^2\} = N_0$. We assume that the system has an average power
constraint of $E\{|s_{x_k}|^2\} \ \le \E \quad \forall k$.

At the transmitter, $M$-ary PSK modulation is employed for
transmission. Hence, the discrete input, $x_k$, takes values from
$\{0,1, \ldots, M-1\}$, and if $x_k = m$, then the transmitted
signal in the $k^{\text{th}}$ symbol duration is
\begin{gather}
s_{x_k} = s_m = \sqrt{\E}e^{j \theta_{m}}
\end{gather}
where $\theta_{m} = \frac{2\pi m}{M} \quad m = 0,1, \ldots, M-1,
$
is one of the $M$ phases available in the constellation.

At the receiver, the detector makes hard decisions for every
received symbol. Therefore, each received signal $r_k$ is mapped to
one of the points in the constellation set $\{\sqrt{\E}e^{j2\pi
m/M}, m = 0,1, \ldots, M-1\}$ before the decoding step. We assume
that maximum likelihood decision rule is used at the detector. Note
that with hard-decision detection, the channel can be now regarded
as a symmetric discrete channel with $M$ inputs and $M$ outputs.

\section{PSK over AWGN Channels} \label{sec:awgn}

We first consider the unfaded AWGN channel and assume that $h = 1$.
In this case, the conditional probability density function of the
channel output given the channel input is\footnote{Since the channel
is memoryless, we henceforth, without loss of generality, drop the
time index $k$ in the equations for the sake of simplification.}
\begin{align}
f_{r|x}(r|x = m) &= f_{r|s_m}(r | s_m)
\\
&=\frac{1}{\pi N_0} e^{-\frac{|r - s_m|^2}{N_0}} \quad m = 0,
\ldots, M-1. \label{eq:condprobr}
\end{align}
It is well-known that the maximum likelihood detector selects the
constellation point closest to the received signal $r$. We denote
the signal at the output of the detector by $y$ and assume that $ y
\in \{0,1, \ldots, M-1\}. $ Note that $y = l$ for $l = 0,1, \ldots,
M-1$ means that the detected signal is $\sqrt{\E} e^{j2\pi l /M}$.
The decision region for $y = l$ is the two-dimensional region
\begin{gather}
D_l = \left\{r = |r| e^{j \theta}:  \frac{(2l-1)\pi}{M} \le \theta <
\frac{(2l+1)\pi}{M} \right\}.
\end{gather}
With quantization at the receiver, the resulting channel is a
symmetric, discrete, memoryless channel with input $x \in \{0,1,
\ldots, M-1\}$ and output $y \in \{0,1, \ldots, M-1\}$. The
transition probabilities are given by
\begin{align}
P_{l,m}  &= P(y = l | x = m)
\\
&=P\left(\frac{(2l-1)\pi}{M} \le \theta < \frac{(2l+1)\pi}{M} | x =
m\right)
\\
&= \int_{\frac{(2l-1)\pi}{M}}^\frac{(2l+1)\pi}{M} f_{\theta |
s_m}(\theta | s_m) \,d \theta
\end{align}
where $f_{\theta | s_m}(\theta | s_m)$ is the conditional
probability density function of the phase of the received signal
given that the input is $x = m$, and hence the transmitted signal is
$s_m$. It is well-known that the capacity of this symmetric channel
is achieved by equiprobable inputs and the resulting capacity
expression \cite{Cover} is
\begin{align}
C_M(\tsnr) &= \log M - H(y|x = 0)
\\
&= \log M + \sum_{l = 0}^{M-1} P_{l,0} \log P_{l,0} \label{eq:mcap}
\end{align}
where $\tsnr = \frac{\E}{N_0}$, $H(\cdot)$ is the entropy function,
and $P_{l,0} = P(y = l | x = 0)$. In order to evaluate the capacity
of general $M$-ary PSK transmission with a hard-decision detector,
the transition probabilities $\{P_{l,0}\}$
should be computed. Starting from (\ref{eq:condprobr}), we can
easily find that
\begin{align}
f_{\theta|s_0}(\theta|s_0) = \frac{1}{2 \pi} \, e^{-\tsnr} +
&\sqrt{\frac{\tsnr}{\pi}} \cos \theta \,e^{-\tsnr \sin^2 \theta}
\nonumber
\\
&\times \left( 1 - Q(\sqrt{2\tsnr \cos^2 \theta})\right)
\label{eq:condprobtheta}
\end{align}
where
\begin{gather} Q(x) = \int_{x}^\infty \frac{1}{\sqrt{2\pi}}
\,e^{-t^2/2} \, dt. \label{eq:qfunction}
\end{gather}
Since $f_{\theta|s_0}$ is rather complicated, closed-form
expressions for the capacity is available only for the special cases
of $M = 2$ and 4:
\begin{align}
C_2(\tsnr) &= \log 2 - h(Q(\sqrt{2\tsnr})) \text{ and } C_4(\tsnr) =
2C_2\left(\frac{\tsnr}{2}\right) \nonumber
\end{align}
where $ h(x) = - x \log x - (1-x)\log(1-x)$. For the other cases,
the channel capacity can only be found by numerical computation.


On the other hand, the behavior of the capacity in the low-$\tsnr$
regime can be accurately predicted through the second-order Taylor
series expansion of the capacity, which involves $\dot{C}_M(0)$ and
$\ddot{C}_M(0)$, the first and second derivatives of the channel
capacity (in nats/symbol) with respect to $\tsnr$ at $\tsnr = 0$. In
the following result, we provide closed-form expressions for these
derivatives.

\begin{theo:awgnderivatives} \label{theo:awgnderivatives}
The first and second derivatives of $C_M(\tsnr)$ in nats per symbol
at $\tsnr = 0$ are given by
\begin{align}
\dot{C}_M(0) = \left\{
\begin{array}{ll}
\frac{2}{\pi} & M = 2
\\
\frac{M^2}{4\pi} \sin^2 \frac{\pi}{M} & M \ge 3
\end{array} \right., \label{eq:firstderivative}
\end{align}
and
\begin{align}
\ddot{C}_M(0) = \left\{
\begin{array}{ll}
\frac{8}{3\pi}\left( \frac{1}{\pi} -1\right) & M = 2
\\
\infty & M = 3
\\
\frac{4}{3\pi}\left( \frac{1}{\pi} -1\right) & M = 4
\\
\psi(M) & M \ge 5
\end{array} \right. \label{eq:secondderivative}
\end{align}
respectively, where
\begin{align}
\psi(M) = \frac{M^2}{16\pi^2} &\left( (2-\pi)\sin^2 \frac{2\pi}{M} +
(M^2 - 4\pi)\sin^4 \frac{\pi}{M}\right. \nonumber
\\
&\,\,\,\,\,\,- \left.2M \sin^2
\frac{\pi}{M}\sin\frac{2\pi}{M}\right). \label{eq:psi}
\end{align}
\end{theo:awgnderivatives}

\emph{Proof}: The main approach is to obtain $\dot{C}_M(0)$ and
$\ddot{C}_M(0)$ by first finding the derivatives of the transition
probabilities $\{P_{l,0}\}$. This can be accomplished by finding the
first and second derivatives of $f_{\theta|s_0}$ with respect to
$\tsnr$. However, the presence of $\sqrt{\tsnr}$ in second part of
(\ref{eq:condprobtheta}) complicates this approach because
$\left.\frac{df_{\theta|s_0}}{d\ssnr}\right|_{\ssnr = 0} = \infty.$
In order to circumvent this problem, we define the new variable $a =
\sqrt{\tsnr}$ and consider
\begin{align}
f_{\theta|s_0}(\theta|s_0) = \frac{1}{2 \pi} \, e^{-a^2} +
&\frac{a}{\sqrt{\pi}} \cos \theta \,e^{-a^2 \sin^2 \theta} \nonumber
\\
&\times \left( 1 - Q(\sqrt{2a^2 \cos^2 \theta})\right).
\label{eq:condprobthetaa}
\end{align}
The following can be easily verified.
\begin{align*}
&f_{\theta|s_0}(\theta|s_0)|_{a = 0} = \frac{1}{2\pi}, \qquad
\left.\frac{df_{\theta|s_0}}{da}\right|_{a = 0} = \frac{\cos
\theta}{2 \sqrt{\pi}},
\\
&\left.\frac{d^2f_{\theta|s_0}}{da^2}\right|_{a = 0} = \frac{\cos 2
\theta}{\pi}, \qquad \left.\frac{df^3_{\theta|s_0}}{da^3}\right|_{a
= 0} = \frac{-3 \cos \theta \sin^2 \theta}{\sqrt{\pi}},
\\
&\left.\frac{df^4_{\theta|s_0}}{da^4}\right|_{a = 0} = \frac{6
\cos^2 2\theta}{\pi} - \frac{8 \cos^4 \theta}{\pi}.
\end{align*}
Using the above derivatives, we can find the first through fourth
derivatives of $P_{l,0}$ with respect to $a$ at $a = 0$. Using the
derivatives of $P_{l,0}$ and performing several algebraic
operations, we arrive to the following Taylor expansion for $C_M(a)$
at $a = 0$:
\begin{align}
\!\!\!\!\!\!\!C_M(a) &= \phi_1(M) \,a^2 + \phi_2(M) \, a^3 +
\phi_3(M) \, a^4 + o(a^4) \label{eq:expansion1}
\\
&=\phi_1(M) \tsnr + \phi_2(M) \tsnr^{3/2} + \phi_3(M) \tsnr^2 +
o(\tsnr^2) \label{eq:expansion2}
\end{align}
where (\ref{eq:expansion2}) follows due to the fact that $a =
\sqrt{\tsnr}$. In the above expansion,
\begin{align}
&\phi_1(M) = \frac{M}{2\pi} \sin^2 \frac{\pi}{M} \sum_{i=1}^M \cos^2
\frac{2\pi i}{M},
\\
&\phi_2(M) = \frac{M}{\pi\sqrt{\pi}} \left(\sin \frac{\pi}{M} \sin
\frac{2\pi}{M} - \frac{M}{6} \sin^3 \frac{\pi}{M}\right)
\sum_{i=1}^M \cos^3 \frac{2\pi i}{M},
\end{align}
and
\begin{align}
&\phi_3(M) = -\frac{M^2}{16\pi} \sin^2 \frac{2\pi}{M} +
\frac{M(\pi+2)}{16\pi^2}\sin^2\frac{2\pi}{M} \sum_{i=1}^M \cos^2
\frac{4\pi i}{M} \nonumber
\\
&\!\!\!\!\!\!\!\!+\left( \left(\frac{M^3}{12\pi^2} -
\frac{M}{3\pi}\right)\sin^4 \frac{\pi}{M} - \frac{M^2}{2\pi^2}
\sin^2 \frac{\pi}{M} \sin \frac{2\pi}{M}\right) \sum_{i=1}^M \cos^4
\frac{2\pi i}{M} \nonumber
\\
&+\frac{M^2}{4 \pi^2} \sin^2 \frac{\pi}{M} \sin \frac{2\pi}{M}
\sum_{i=1}^M \cos^2 \frac{2\pi i}{M}.
\end{align}
We immediately conclude from (\ref{eq:expansion2}) that
$\dot{C}_M(0) = \phi_1(M)$. Note that the expansion includes the
term $\tsnr^{3/2}$ which implies that $\ddot{C}_M(0) = \pm\infty$
for all $M$. However, it can be easily seen that $\phi_2(M) = 0$ for
all $M \neq 3$, and at $M = 3$, $\phi_2(3) = 0.1718$. Therefore,
while $\ddot{C}_3(0) = \infty$, $\ddot{C}_M(0) = 2\phi_3(M)$ for $M
\neq 3$. Further algebraic steps and simplification yields
(\ref{eq:firstderivative}) and (\ref{eq:secondderivative}). \hfill
$\square$

\emph{Remark:} We should note that the first derivative expression
(\ref{eq:firstderivative}) has previously been given in
\cite{Pierce} through the bit energy expressions. In addition,
Verd\'u in \cite{Verdu} has provided the second derivative
expression for the special case of $M = 4$. Hence, the main novelty
in Theorem \ref{theo:awgnderivatives} is the second derivative
expression for general $M$. First and second derivative expressions
are given together for completeness.

The following corollary provides the asymptotic behavior as $M \to
\infty$.

\begin{corr:awgnasympt}
In the limit as $M \to \infty$, the first and second derivatives of
the capacity at zero $\tsnr$ converge to
\begin{align}
\lim_{M\to \infty} \dot{C}_M(0) = \frac{\pi}{4} \text{ and } \lim_{M
\to \infty} \ddot{C}_M(0) = \frac{\pi^2-8\pi+8}{16}.
\end{align}
\end{corr:awgnasympt}\vspace{.2cm}

In the low-power regime, the tradeoff between bit energy and
spectral efficiency is a key measure of performance. The normalized
energy per bit can be obtained from
\begin{gather}
\frac{E_b}{N_0} = \frac{\tsnr}{C(\tsnr)}
\end{gather}
where $C(\tsnr)$ is the channel capacity in bits/symbol. The maximum
achievable spectral efficiency in bits/s/Hz is given by
\begin{gather}
\C\left(\frac{E_b}{N_0} \right) = C(\tsnr) \text{ bits/s/Hz}
\end{gather}
if we, without loss of generality, assume that one symbol occupies a
1s $\times$ 1Hz time-frequency slot. Two important notions regarding
the spectral-efficiency/bit-energy tradeoff in the low power regime
are the bit-energy required at zero spectral efficiency,
\begin{gather}
\left.\frac{E_b}{N_0}\right|_{\C = 0} = \frac{\log_e2}{\dot{C}(0)},
\end{gather}
and the wideband slope,
\begin{gather}
S_0 = \frac{2 (\dot{C}(0))^2}{-\ddot{C}(0)},
\end{gather}
which gives the slope of the spectral efficiency curve $\C(E_b/N_0)$
at zero spectral efficiency \cite{Verdu}. Therefore,
$\left.\frac{E_b}{N_0}\right|_{\C = 0}$ and $S_0$ constitute a
linear approximation to the spectral efficiency curve in the
low-$\tsnr$ regime. Since these quantities depend only $\dot{C}(0)$
and $\ddot{C}(0)$, the bit energy at zero spectral efficiency and
wideband slope achieved by $M$-ary PSK signals with a hard-decision
detector can be readily obtained by using the formulas
(\ref{eq:firstderivative}) and (\ref{eq:secondderivative}).
\begin{corr:awgnbitenergy}
The bit energy at zero spectral efficiency and wideband slope
achieved by $M$-ary PSK signaling are given by
\begin{align}
\left.\frac{E_b}{N_0}\right|_{\C = 0} = \left\{
\begin{array}{ll}
\frac{\pi}{2}\log_e2 & M = 2
\\
\frac{4 \pi}{M^2 \sin^2 \frac{\pi}{M}}\log_e 2 & M \ge 3
\end{array}\right.
\end{align}
and
\begin{align}
S_0 = \left\{
\begin{array}{ll}
\frac{3}{\pi-1}& M = 2
\\
0 & M = 3
\\
\frac{6}{\pi-1} & M = 4
\\
\frac{\frac{M^4}{8\pi^2}\sin^4 \frac{\pi}{M}}{-\psi(M)} & M \ge 5
\end{array} \right. 
\end{align}
respectively.
\end{corr:awgnbitenergy}

As it will be evident in numerical results, generally the
$\left.\frac{E_b}{N_0}\right|_{\C = 0}$ is the minimum bit energy
required for reliable transmission when $M \neq 3$. However, for $M
= 3$, the minimum bit energy is achieved at a nonzero spectral
efficiency.

\begin{corr:awgnM3bitenergy}
For 3-PSK modulation, the minimum bit energy is achieved at a
nonzero spectral efficiency.
\end{corr:awgnM3bitenergy}

This corollary follows immediately from the fact that $\ddot{C}_3(0)
= \infty$ which implies that the slope at zero $\tsnr $ of
$\tsnr/C_3(\tsnr)$ is $-\infty$. This lets us conclude that the bit
energy required at zero spectral efficiency cannot be the minimum
one. The fact that 3-PSK achieves its minimum bit energy at a
nonzero spectral efficiency is also pointed out in \cite{Kramer}
through numerical results. Here, this result is shown analytically
through the second derivative expression.

\begin{figure}
\begin{center}
\includegraphics[width = 0.45\textwidth]{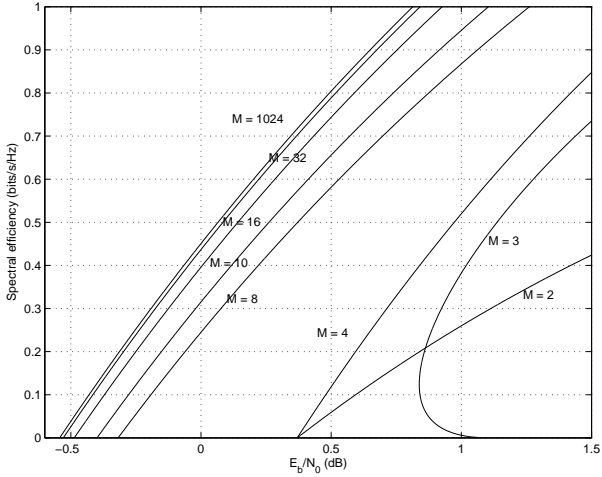}
\caption{Spectral efficiency $\C(E_b/N_0)$ vs. bit energy $E_b/N_0$
for $M$-ary PSK with a hard-decision detection in the AWGN channel.}
\label{fig:awgnbitenergy}
\end{center}
\end{figure}

Figure \ref{fig:awgnbitenergy} plots the spectral efficiency curves
as a function of the bit energy for hard-detected PSK with different
constellation sizes. As observed in this figure, the
information-theoretic analysis conducted in this paper provides
several practical design guidelines. We note that although 2-PSK and
4-PSK achieve the same minimum bit energy of 0.369 dB at zero
spectral efficiency, 4-PSK is more efficient at low but nonzero
spectral efficiency values due to its wideband slope being twice
that of 2-PSK. 3-PSK is better than 2-PSK for spectral efficiency
values greater than 0.208 bits/s/Hz below which 3-PSK performes
worse than both 2 and 4-PSK. 3-PSK achieves its minimum bit energy
of 0.8383 dB at 0.124 bits/s/Hz. Operation below this level of
spectral efficiency should be avoided as it only increases the
energy requirements. We further observe that increasing the
constellation size to 8 provides much improvement over 4-PSK. 8-PSK
achieves a minimum bit energy of $-0.318$ dB. Further increase in
$M$ provides diminishing returns. For instance, there is little to
be gained by increasing the constellation size more than 32 as
32-PSK achieves a minimum bit energy of $-0.528$ dB and the minimum
bit energy as $M \to \infty$ is $-0.542$ dB. Note that $-0.542$ dB
still presents a loss of approximately 1.05 dB with respect to the
fundamental limit of $-1.59$ dB achieved by soft detection. We find
that the wideband slopes of M = 8,10,16,32, and 1024 are 2.44, 2.53,
2.64, 2.69, and 2.71. The similarity of the wideband slope values is
also apparent in the figure. As $M \to \infty$, the wideband slope
is 2.717. Finally, note that the wideband slope of 3-PSK, as
predicted, is 0.

\section{PSK over Fading Channels}

\subsection{Coherent Fading Channels}

In this section, we consider fading channels and assume that the
fading coefficients $\{h_k\}$ are known at the receiver but not at
the transmitter. The only requirements on the fading coefficients
are that their variations are ergodic and they have finite second
moments. Due to the presence of receiver channel side information
(CSI), scaled nearest point detection is employed, and the analysis
follows along lines similar to those in the previous section. Hence,
the treatment will be brief.

Note that in this case, the average capacity is
\begin{gather}
C_{M}(\tsnr) = \log M + \sum_{l = 0}^{M-1} E_h \{P_{l,0,h} \log
P_{l,0,h}\}
\end{gather}
where
\begin{gather}
P_{l,0,h} = \int_{\frac{(2l-1)\pi}{M}}^\frac{(2l+1)\pi}{M} f_{\theta
| s_0,h}(\theta | s_0,h) \,d \theta
\end{gather}
and
\begin{align}
f_{\theta|s_0,h}(\theta|s_0,h) = \frac{1}{2 \pi} &\, e^{-|h|^2\tsnr}
+ \sqrt{\frac{|h|^2\tsnr}{\pi}} \cos \theta \,e^{-|h|^2\tsnr \sin^2
\theta} \nonumber
\\
&\times \left( 1 - Q(\sqrt{2|h|^2\tsnr \cos^2 \theta})\right)
\label{eq:cfcondprobtheta}
\end{align}
with $\tsnr = \E/N_0$. Through a similar analysis as in Section
\ref{sec:awgn}, we have the following result on the derivatives of
the capacity.
\begin{theo:cfderivatives} \label{theo:cfderivatives}
The first and second derivatives of $C_M(\tsnr)$ in nats per symbol
at $\tsnr = 0$ are given by
\begin{align}
\dot{C}_M(0) = \left\{
\begin{array}{ll}
\frac{2}{\pi}E\{|h|^2\} & M = 2
\\
\frac{M^2}{4\pi} \sin^2 \frac{\pi}{M} E\{|h|^2\}& M \ge 3
\end{array} \right., \label{eq:cfadingfirstderivative}
\end{align}
and
\begin{align}
\ddot{C}_M(0) = \left\{
\begin{array}{ll}
\frac{8}{3\pi}\left( \frac{1}{\pi} -1\right)E\{|h|^4\} & M = 2
\\
\infty & M = 3
\\
\frac{4}{3\pi}\left( \frac{1}{\pi} -1\right) E\{|h|^4\} & M = 4
\\
\psi(M) E\{|h|^4\} & M \ge 5
\end{array} \right. \label{eq:cfadingsecondderivative}
\end{align}
respectively, where $\psi(M)$ is given in (\ref{eq:psi}).
\end{theo:cfderivatives}

Note that the first derivative and second derivatives of the
capacity at zero $\tsnr$ are essentially equal to the scaled
versions of those obtained in the AWGN channel. The scale factors
are $E\{|h|^2\}$ and $E\{|h|^4\}$ for the first and second
derivatives, respectively.

In the fading case, we can define the received bit energy as
\begin{gather}
\frac{E_b^r}{N_0} = \frac{E\{|h|^2\}\tsnr}{C_M(\tsnr)}
\end{gather}
as $E\{|h|^2\} \tsnr$ is the received signal-to-noise ratio. It
immediately follows from Theorem \ref{theo:cfderivatives} that
$E_b^r / N_0 |_{\C = 0}$ in the coherent fading channel is the same
as that in the AWGN channel. On the other hand, the wideband slope
is scaled by $(E\{|h|^2\})^2/E\{|h|^4\}$.

\subsection{Noncoherent Fading Channels}

In this section, we assume that neither the receiver nor the
transmitter knows the fading coefficients $\{h_k\}$. We further
assume that $\{h_k\}$ are i.i.d. proper complex Gaussian random
variables with mean $E\{h_k\} = d \neq 0$ \footnote{$d \neq 0$ is
required because phase cannot be used to transmit information in a
noncoherent Rayleigh fading channel where $d = 0$.} and variance
$E\{|h_k - d|^2\} = \gamma^2$. Now, the conditional probability
density function of the channel output given the input is
\begin{align}
f_{r|s_m}(r | s_m) =\frac{1}{\pi (\gamma^2|s_m|^2 + N_0)}
e^{-\frac{|r - ds_m|^2}{\gamma^2|s_m|^2 + N_0}}. 
\label{eq:nfcondprobr}
\end{align}
Recall that $\{s_m = \sqrt{\E} e^{j\theta_m}\}$ are the PSK signals
and hence $|s_m| = \sqrt{\E}$ for all $m = 0, \ldots, M-1$. Due to
this constant magnitude property, it can be easily shown that the
maximum likelihood detector selects $s_k$ as the transmitted signal
if\footnote{(\ref{eq:decisionrule}) is obtained when we assume,
without loss of generality, that $d = |d|$.}
\begin{gather}\label{eq:decisionrule}
\text{Re}(r s_k^*) > \text{Re}(r s_i^*) \quad \forall i \neq k
\end{gather}
where $s_k^*$ is the complex conjugate of $s_k$, and $\text{Re}$
denotes the operation that selects the real part. Therefore, the
decision regions are the same as in the AWGN channel case.

In this case, the channel capacity is
\begin{gather}
C_{M}(\tsnr) = \log M + \sum_{l = 0}^{M-1} P_{l,0} \log P_{l,0}
\end{gather}
where \vspace{-.5cm}
\begin{gather}
P_{l,0} = \int_{\frac{(2l-1)\pi}{M}}^\frac{(2l+1)\pi}{M} f_{\theta |
s_0}(\theta | s_0) \,d \theta
\end{gather}
and
\begin{align}
f_{\theta|s_0}(\theta|s_0) = &\frac{1}{2 \pi} \,
e^{-\frac{|d|^2\tsnr}{\gamma^2 \tsnr + 1}}
\\+
&\sqrt{\frac{|d|^2\tsnr}{\pi(\gamma^2 \tsnr + 1)}} \cos \theta
\,e^{-\frac{|d|^2\tsnr}{\gamma^2 \tsnr + 1} \sin^2 \theta} \nonumber
\\
&\times \left( 1 - Q\left(\sqrt{2\frac{|d|^2\tsnr}{\gamma^2 \tsnr +
1} \cos^2 \theta}\right)\right). \label{eq:nfcondprobtheta}
\end{align}

The following results provide the first and second derivatives of
the capacity at zero $\tsnr$, and the bit energy and wideband slope
in the low-$\tsnr$ regime.

\begin{theo:ncfderivatives}
The first and second derivatives of $C_M(\tsnr)$ in nats per symbol
at $\tsnr = 0$ are given by
\begin{align}
\dot{C}_M(0) = \left\{
\begin{array}{ll}
\frac{2|d|^2}{\pi} & M = 2
\\
\frac{M^2|d|^2}{4\pi} \sin^2 \frac{\pi}{M} & M \ge 3
\end{array} \right., \label{eq:ncfadingfirstderivative}
\end{align}
and
\begin{align}
\ddot{C}_M(0) = \left\{
\begin{array}{ll}
\frac{8}{3\pi}\left( \frac{1}{\pi} -1\right)|d|^4 - \frac{4 |d|^2
\gamma^2}{\pi} & M = 2
\\
\infty & M = 3
\\
\frac{4}{3\pi}\left( \frac{1}{\pi} -1\right)|d|^4 - \frac{4 |d|^2
\gamma^2}{\pi} & M = 4
\\
\psi(M)|d|^4-\frac{|d|^2 \gamma^2}{2\pi}M^2 \sin^2\frac{\pi}{M} & M
\ge 5
\end{array} \right. \label{eq:ncfadingsecondderivative}
\end{align}
respectively, where $\psi(M)$ is given in (\ref{eq:psi}).
\end{theo:ncfderivatives}

\begin{corr:ncfadingasympt}
In the limit as $M \to \infty$, the first and second derivatives of
the capacity at zero $\tsnr$ converge to
\begin{gather}
\lim_{M\to \infty} \dot{C}_M(0) = \frac{\pi|d|^2}{4}.
\end{gather}
 \vspace{-.3cm}
and
\begin{gather}
 \lim_{M \to \infty} \ddot{C}_M(0) =
\frac{(\pi^2-8\pi+8)|d|^4}{16}-\frac{|d|^2 \gamma^2 \pi}{2}.
\end{gather}
\end{corr:ncfadingasympt}

In the noncoherent fading case, the received bit energy is
\begin{gather}
\frac{E_b^r}{N_0} = \frac{(|d|^2 + \gamma^2)\tsnr}{C_M(\tsnr)}.
\end{gather}

\begin{corr:ncfadingbitenergy}
The received bit energy at zero spectral efficiency and wideband
slope achieved by $M$-ary PSK signaling are given by
\begin{align}
\left.\frac{E_b}{N_0}\right|_{\C = 0} = \left\{
\begin{array}{ll}
\frac{\pi}{2}\left( 1 + \frac{1}{\K}\right)\log_e2 & M = 2
\\
\frac{4 \pi}{M^2 \sin^2 \frac{\pi}{M}} \left( 1 +
\frac{1}{\K}\right)\log_e 2 & M \ge 3
\end{array}\right.
\end{align}
and
\begin{align}
S_0 = \left\{
\begin{array}{ll}
\frac{3}{\pi-1 + \frac{3\pi}{2\K}}& M = 2
\\
0 & M = 3
\\
\frac{6}{\pi-1 + \frac{3\pi}{\K}} & M = 4
\\
\frac{\frac{M^4}{8\pi^2}\sin^4 \frac{\pi}{M}}{-\psi(M) +
\frac{1}{2\pi \K}M^2 \sin^2\frac{\pi}{M}} & M \ge 5
\end{array} \right. \label{eq:ncfadingsecondderivative}
\end{align}
respectively, where $\psi(M)$ is given in (\ref{eq:psi}), and $\K =
\frac{|d|^2}{\gamma^2}$ is the Rician factor.
\end{corr:ncfadingbitenergy}

\emph{Remark}: If we let $|d| = 1$ and $\gamma^ 2 = 0$, or
equivalently let $\K \to \infty$, the results provided above
coincide with those given for the AWGN channel.

\begin{figure}
\begin{center}
\includegraphics[width = 0.45\textwidth]{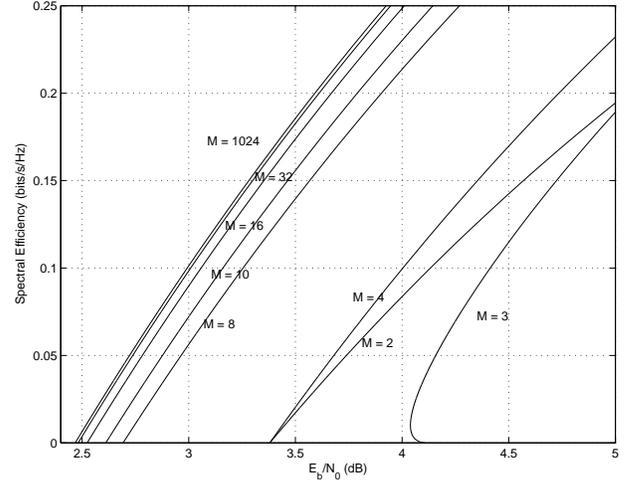}
\caption{Spectral efficiency $\C(E_b/N_0)$ vs. bit energy $E_b/N_0$
in the noncoherent Rician fading channel with Rician factor $\K =
\frac{|d|^2}{\gamma^2} = 1$.} \label{fig:ncfadingbitenergy}
\end{center}
\end{figure}

Fig. \ref{fig:ncfadingbitenergy} plots the spectral efficiency
curves as a function of the bit energy for $M$-ary PSK transmission
over the noncoherent Rician fading channel with $\K = 1$. Note that
conclusions similar to those given for Fig. \ref{fig:awgnbitenergy}
also apply for Fig. \ref{fig:ncfadingbitenergy}. The main difference
between the figures is the substantial increase in the bit energy
values as a penalty of not knowing the channel. For instance, 2 and
4-PSK now achieves a minimum bit energy of 3.379 dB while 8-PSK
attains 2.692 dB. As $M \to \infty$, the minimum bit energy goes to
2.467 dB.


%


\begin{thebibliography}{}

\bibitem{Pierce} J. R. Pierce, ``Comparison of three-phase
modulation with two-phase and four-phase modulation," \emph{IEEE
Trans. Commun}, vol.~28, pp.~1098-1099, July 1980.

\bibitem{Kramer} G. Kramer, A. Ashikhmin, A. J. van Wijngaarden, and X. Wei, ``Spectral efficiency of coded
phase-shift keying for fiber-optic communication," \emph{IEEE/OSA J.
Lightwave Technol.}, vol.~21, pp.~2438-2445, Oct. 2003.

\bibitem{Verdu} S. Verd\'u, ``Spectral efficiency in the wideband regime," \emph{IEEE
Trans. Inform. Theory}, vol.~48, pp.~1319-1343, June 2002.

\bibitem{Wyner} A. D. Wyner, ``Bounds on communication with
polyphase coding," \emph{Bell Syst. Tech. J.,} vol. XLV, pp.
523-559, Apr. 1966.

\bibitem{Kaplan} G. Kaplan and S. Shamai (Shitz), ``On the
achievable information rates of DPSK," \emph{IEE Proceesings}, vol.
139, pp. 311-318, June 1992.

\bibitem{Peleg} M. Peleg and Shlomo Shamai (Shitz), ``On the
capacity of the blockwise incoherent MPSK channel," \emph{IEEE
Trans. Commun}, vol.~46, pp.~603-609, May 1998.

\bibitem{Gursoy-part2} M. C. Gursoy, H. V. Poor, and S. Verd\'u, ``The noncoherent Rician fading
channel -- Part II : Spectral efficiency in the low power regime,"
\emph{IEEE Trans. Wireless Commun.}, vol. 4, no. 5, pp. 2207-2221,
Sept. 2005.

\bibitem{Zhang} W. Zhang and J. N. Laneman,``How good is phase-shift keying for
peak-limited Rayleigh fading channels in the low-SNR regime?," to
appear in \emph{IEEE Trans. Inform. Theory}.

\bibitem{Luo} X. Luo and G. B. Giannakis, ``Energy-constrained optimal
quantization for wireless sensor networks," IEEE SECON, pp. 272-278,
4-7 Oct. 2004.


\bibitem{Proakis} J. G. Proakis, \emph{Digital Communications.}
New York: McGraw-Hill, 1995.

\bibitem{Cover} T. M. Cover and J. A. Thomas, \emph{Elements of Information
Theory.} New York: Wiley, 1991.

\end{thebibliography}
\end{document}